\documentclass[12pt, a4paper]{article} %
\usepackage{graphicx} %
\usepackage[utf8]{inputenc} %
\usepackage{color} %
\usepackage{listings} %
\usepackage{hyperref}
\hypersetup{
    colorlinks,%
    citecolor=black,%
    filecolor=black,%
    linkcolor=black,%
    urlcolor=black,
	pdftitle=Payez I (brief report),pdfdisplaydoctitle
}
\usepackage{amsfonts} %
\usepackage{amsmath} %
\usepackage{amssymb} %
\usepackage{mathrsfs} %
\usepackage{latexsym} %
\usepackage{euscript} %
\usepackage{cite} %[1,2,3,4] >> [1-4]
\usepackage{mciteplus} %
\DeclareTextSymbol{\degre}{OT1}{23}

\newcommand \be{\begin{equation}} %
\newcommand \ee{\end{equation}} %
\newcommand \bea{\begin{eqnarray}} %
\newcommand \eea{\end{eqnarray}} %
\newcommand \ba{\begin{array}} %
\newcommand \ea{\end{array}} %

\title{
	\textsc{
		Cornering the axion-like particle explanation of quasar polarisations
	}
}

\author{A.~Payez\thanks{e-mail: A.Payez@ulg.ac.be.}\\\footnotesize{\emph{IFPA group, AGO Dept., U.~of Li\`ege, B4000 Li\`ege, Belgium}}
}

\date{
}

\begin{document}

	\maketitle

	\begin{abstract}
		In a series of paper, it has been shown that the distribution of polarisation position angles for visible light from quasars is not random in extremely large regions of the sky. As explained in a recent article, the measurement of vanishing circular polarisation for such quasars is an important problem for a mechanism involving the mixing with axion-like particles in external magnetic fields. In this note, we stress that a recent report of similar coherent orientations of polarisation in radiowaves further disfavours the need for such particles, as an effect at these wavelengths would be extremely suppressed or would directly contradict data.
	\end{abstract}

		The existence of coherent orientations in the polarisation of visible light from quasars in large-scale ($\sim 1$~Gpc) regions of the sky~\cite{Hutsemekers:1998, *Hutsemekers:2001, *Hutsemekers:2005,Jain:2003sg} is a very puzzling observation. To date, no satisfactory explanation of this effect, based on 355 high-quality measurements of linear polarisation from quasars, is available.

		Until recently, the best hope to explain these data involved axion-like particles (ALPs), extremely light spinless particles with a coupling to two photons similar to the Primakoff effect for neutral pions.
		In a nutshell, light from these quasars could mix with such ALPs inside the external magnetic fields\footnote{Let us stress that this work only deals with the mixing of light with ALPs inside external magnetic fields. In particular, the case of the propagation of light inside an hypothetical anisotropic scalar background field, for which the phenomenology is very different~\cite{Harari:1992ea}, is beyond the scope of this note.} encountered on their way towards Earth~\cite{Jain:2002vx,*Das:2004qka,*Piotrovich:2008iy,*Payez:2008pm,*Hutsemekers:2008iv,*Payez:2009kc,*Payez:2009vi,*Payez:2010xb,*Agarwal:2009ic}. Given the dichroic property of the mixing, it was believed that coherent alignments could be reproduced from initial random distributions of polarisation position angles.

		However, in a recent publication~\cite{Payez:2011sh}, we have shown that measurements of null circular polarisation in V-filter for objects taken from the 355-quasar sample~\cite{Hutsemekers:2010fw}  is in severe contradiction with the phenomenology expected from ALP-photon mixing. The production of alignments of linear polarisation, while keeping the circular polarisation small, fails already at the qualitative level\hspace{1pt}---\hspace{1pt}even within a wave-packet treatment, including fluctuations, and considering more refined magnetic field configurations. As there is either too much circular polarisation, too much linear polarisation, or no alignment in general, we concluded in~\cite{Payez:2011sh} that this mechanism was strongly disfavoured by data.

		Recently, a very interesting data analysis~\cite{Tiwari:2012rr} has suggested that similar coherent orientations in the linear polarisation of quasars also exist in radio wavelengths (8.4~GHz). While the authors do agree with the original data analysis~\cite{Joshi:2007yf} in the restricted case which was considered at the time,\footnote{The original study was limited to the case of 4290 quasar polarisation measurements (52 of these objects being part of the 355-quasar sample of optical measurements) and detected no alignment.} they also report very significant evidences of alignments (up to 5$\sigma$) for different data cuts. What is also extremely interesting is that these alignments have been detected using the same coordinate-invariant statistics one of the authors used for alignments in visible light in~\cite{Jain:2003sg}.

			If there are similar large-scale effects in different energy domains, it is quite natural to think that they have the same origin.
			In the following, we emphasise that such an observation of alignments in radio wavelengths cannot be explained by ALP-photon mixing in external magnetic fields. On the one hand, this is not expected in the (already excluded) scenario associated with alignments in visible light; on the other hand, if the mixing was efficient enough in radiowaves, then there would be a strong contradiction with polarisation data in visible light.

			This can be shown starting from the Lagrangian density:\footnote{Note that, henceforth, we particularise to the pseudoscalar case but similar results can be derived for scalar ALPs.}
			\begin{equation}
				 \mathcal{L} =  \frac{1}{2}\ (\partial_{\mu}\phi) (\partial^{\mu}\phi) - \frac{1}{2}\ m^2\phi^2 - \frac{1}{4}\ F_{\mu\nu} F^{\mu\nu} + \frac{1}{4}\ g \phi F_{\mu\nu}\widetilde{F}^{\mu\nu}, \label{eq:lagrangian}
			\end{equation}
			where $\widetilde{F}^{\mu\nu}\equiv\frac{1}{2}\ \epsilon^{\mu\nu\rho\sigma}F_{\rho\sigma}$ is the dual of the electromagnetic tensor, $m$ is the ALP mass and $g$ is the coupling constant of the interaction between ALPs and photons. In a magnetic field region, one can then derive that the maximum amount of polarisation attainable for a light beam of frequency $\omega$ is entirely determined by the mixing angle
			\be
				\theta_{\textrm{mix}} = \frac{1}{2} \textrm{atan}\left(\frac{2g\mathcal{B}\omega}{m^2 - {\omega_{\mathrm{p}}}^2}\right);\label{eq:thetamix}
			\ee
			where $\mathcal{B}$ is the strength of the external transverse magnetic field, and $\omega_{\mathrm{p}}$ the plasma frequency.
			In the absence of an initial propagating ALP field $\phi(0)$, the Stokes parameters (the observed physical quantities) of a light beam described initially by $(I_0, Q_0, U_0, V_0)$ will evolve in the following way~\cite{Payez:2011sh}:
		        \bea
		        \left\{
		            \begin{array}{llll}
		                I(z) &=& I_0 - \frac{1}{2}\left(I_0 + Q_0\right) \sin^2 2\theta_{\textrm{mix}} \sin^2\left(\frac{1}{4}\frac{\Delta\mu^2}{\omega}z\right);\\
		                Q(z) &=& I(I_0\longleftrightarrow  Q_0);\\
		                U(z) &=& U_0\big\{  {(\textrm{s}_{\textrm{mix}})}^2 \cos\left(\frac{1}{2}{\left(\textrm{c}_{\textrm{mix}}\right)}^2\ \frac{\Delta\mu^2}{\omega}z\right) + {(\textrm{c}_{\textrm{mix}})}^2 \cos\left(\frac{1}{2}{\left(\textrm{s}_{\textrm{mix}}\right)}^2\ \frac{\Delta\mu^2}{\omega}z\right) \big\}\\
				&& \!\!\!\!- V_0\big\{ {(\textrm{s}_{\textrm{mix}})}^2 \sin\left(\frac{1}{2}{\left(\textrm{c}_{\textrm{mix}}\right)}^2\ \frac{\Delta\mu^2}{\omega}z\right) \ \!- {(\textrm{c}_{\textrm{mix}})}^2 \sin\left(\frac{1}{2}{\left(\textrm{s}_{\textrm{mix}}\right)}^2\ \frac{\Delta\mu^2}{\omega}z\right) \big\} \textrm{sign}(\theta_{\textrm{mix}});\\
		                V(z) &=& U(U_0\rightarrow V_0, V_0\rightarrow-U_0);
		            \end{array} \right.
		            \label{eq:Stokes_alternative}
		        \eea
			where we write $\textrm{c}_{\textrm{mix}}\equiv\cos{\theta_{\textrm{mix}}}$, $\textrm{s}_{\textrm{mix}}\equiv\sin{\theta_{\textrm{mix}}}$, $z$ the distance travelled inside the magnetic field, and where $\Delta\mu^2=\sqrt{{{(2g\mathcal{B}\omega)}^2 + (m^2 - {\omega_{\mathrm{p}}}^2)}^2}$ is the difference of the masses squared of the eigenstates of the mixing. In Eq.~\eqref{eq:Stokes_alternative}, $I$ is the intensity, $Q$ and $U$ describe the linear polarisation, and $V$ is the circular polarisation.
			Now, if one keeps $\phi(0)\neq0$, the evolution of the Stokes parameters is of course more complicated. Nevertheless, the relevant parameters which drive the change of polarisation remain the two dimensionless quantities $\theta_{\textrm{mix}}$ and $\frac{\Delta\mu^2}{\omega}z$, as in the simpler case discussed here.\footnote{Note that, while the mixing angle $\theta_{\textrm{mix}}$ controls the maximum amount of polarisation that can be reached, $\frac{\Delta\mu^2}{\omega}z$ is responsible for the details of the propagation, as in the case of neutrino oscillations.}

			Let us now consider the emission of light from a given quasar, and focus on two values of the frequency: $\omega_1$ and $\omega_2$, which will be redshifted as light propagates. Let us choose $\omega_1$ in such a way that it will be observed as visible light, $\omega_V = 2.5$~eV corresponding to 500~nm; while $\omega_2$ will be redshifted in the radio band, $\omega_R = 3.474\times10^{-5}$~eV corresponding to the observations at 8.4~GHz. For these two beams, the external conditions will be the same as they propagate towards us, so that we have at all times:\footnote{Here, we neglect the tiny difference of group velocity which is formally caused by the non-zero plasma frequency. Equivalently, we can suppose that the external conditions do not change on the time scale which separates the two wave fronts.}
			\be
			\frac{\tan\left(2\theta_{\textrm{mix}}(\omega_2)\right)}{\tan\left(2\theta_{\textrm{mix}}(\omega_1)\right)} = \frac{\omega_R}{\omega_V}=1.4\times10^{-5},\label{eq:compare_theta}
			\ee
			as these beams are redshifted in the same way.
			From Eq.~\eqref{eq:compare_theta} and the discussion above, it is already clear that the effect due to ALP-photon mixing in radiowaves is inefficient compared to the one in visible light.

			As discussed in~\cite{Payez:2011sh}, in order to produce an additional polarisation similar to the one needed in optical wavelengths, $\theta_{\textrm{mix}}(\omega_1)=0.1$ is a typical value.
			Now, to determine the corresponding value for $\omega_2$, we can approximate $\tan\left(x\right)\approx x$ in Eq.~\eqref{eq:compare_theta}. Doing so, we introduce a relative error slightly bigger than 1\% for $\tan\left(2\theta_{\textrm{mix}}(\omega_1)\right)$, but it allows us to continue the discussion in the general case.\footnote{One can also check this result using directly $\omega_V$ and $\omega_R$, and choosing values for the parameters $g$, $m$, $\mathcal{B}$, and $\omega_{\mathrm{p}}$ such that $\theta_{\textrm{mix}}(\omega_V)=0.1$. With the same parameters, one can then calculate $\theta_{\textrm{mix}}(\omega_R)$.}
			We then finally obtain that, while $\theta_{\textrm{mix}}(\omega_1)=0.1$, the mixing angle corresponding to the other light beam is as small as $\theta_{\textrm{mix}}(\omega_2)=1.4\times10^{-6}$ under the same external conditions.

			In order to give a quantitative estimate of the additional polarisation that is typically brought by ALP-photon mixing in both cases, let us consider initially unpolarised light beams in a magnetic field region.
			The degree of linear polarisation evolves in the following way~\cite{Payez:2011sh}:
			\be
				p_{\mathrm{lin}}(z)=\frac{\frac{1}{2}\sin^2 2\theta_{\mathrm{mix}}\sin^2[\frac{1}{4}\frac{\Delta\mu^2}{\omega}z]}
						   {1 - \frac{1}{2}\sin^2 2\theta_{\mathrm{mix}}\sin^2[\frac{1}{4}\frac{\Delta\mu^2}{\omega}z]}.
				\label{eq:plinrtheta}
			\ee
			One can then drop the information associated with the propagation and simply check the maximum amount of linear polarisation that can be achieved in this region, namely:
			\be
				p_{\mathrm{lin}}\big|_{\mathrm{max}}(\omega)=\frac{\frac{1}{2}\sin^2 2\theta_{\mathrm{mix}}(\omega)}
								   {1 - \frac{1}{2}\sin^2 2\theta_{\mathrm{mix}}(\omega)}.
				\label{eq:plinmax}
			\ee
			Finally, we use the values of $\theta_{\mathrm{mix}}$ that we obtained for $\omega_1$ and for $\omega_2$, and replace them in Eq.~\eqref{eq:plinmax}.
			For an additional linear polarisation of $p_{\mathrm{lin}}\big|_{\mathrm{max}}(\omega_1)=2\%$ for what would be visible light, we only have at most a very tiny $p_{\mathrm{lin}}\big|_{\mathrm{max}}(\omega_2)=4\times10^{-10}\%$ in the other case, which is far smaller than what can be detected experimentally.
			The mixing of photons with ALPs in external magnetic fields  thus cannot produce, for the same source, an observable effect in different energy regimes such as visible and radio. Note, of course, that an alignment sufficiently important in radiowaves with this mechanism would imply too much polarisation in visible light, which would contradict the observations: the observed polarisation in visible light is indeed mainly of intrinsic origin~\cite{Hutsemekers:1998,*Hutsemekers:2001,*Hutsemekers:2005,Hutsemekers:1998pp,*Schmidt:1999p,*Lamy:2004yz}.

			While things can become more elaborate in more complex magnetic fields, the phenomenology we have discussed remains the same: an effect in radio wavelengths would be so limited that we should not expect to observe it, according to the scenario in which ALPs provide the mechanism responsible for coherent alignments in visible light. Additionally, other magnetic field configurations will not produce more polarisation than this toy-model in general, as fluctuations tend to diminish the amount created via the mixing.
\bigskip

			In summary, we have shown in this work that ALP-photon mixing in external magnetic fields cannot explain the recent claim of very significant large-scale alignments of quasar polarisations in radio wavelengths.

			This is further evidence that these particles do not explain such alignments of polarisation.
			Indeed, as shown in~\cite{Payez:2011sh}, the existence of similar alignments in visible light cannot be understood through this mechanism either. Until recently, it was thought that ALP-photon mixing might generate large-scale coherent orientations in visible light, provided that there were coherent large-scale magnetic fields;\footnote{Note that one had to justify why this would generate two different preferred directions of polarisation along the same line of sight, depending on the redshift~\cite{Hutsemekers:1998,*Hutsemekers:2001,*Hutsemekers:2005}, which is not trivial.} however, the price to pay for this is the introduction of a circular polarisation problem which directly contradicts high-precision polarisation data.

			It could be that the two very similar observations of large-scale coherent orientations of polarisation of radiowaves and optical light from quasars require completely different physical explanations. Nevertheless, we stress that the ALP mechanism simply fails to reproduce polarisation data both in visible (already at the qualitative level~\cite{Payez:2011sh}) and in radio wavelengths (as it would lead to an extremely efficient mixing in visible light which would contradict data).

			As a side note, some effort has been done in new directions to try to explain the alignments in visible light~\cite{Urban:2011,*Antoniou:2010gw,*Poltis:2010yu,*MosqueraCuesta:2011tz,*Ciarcelluti:2012pc} and some of them could also naturally explain alignments in radiowaves, without generating any circular polarisation.
			While more quantitative predictions are still needed, the new radio data analysis makes these kinds of models quite appealing.
\bigskip

	It is a pleasure to thank Jean-Ren\'e Cudell, Damien Hutsem\'ekers, and Paolo Ciarcelluti for useful discussions and comments about this issue.

\small
\bibliographystyle{modMminimalist}
\bibliography{alexbib}

\end{document}